\documentclass[a4paper,12pt]{article}
\usepackage[english]{babel}
\usepackage{amsmath}
\usepackage{amssymb}
\usepackage{amsfonts}
\usepackage{graphicx}
\usepackage[colorinlistoftodos]{todonotes}
\usepackage[toc,page]{appendix}
\usepackage{geometry}
 \usepackage{hyperref}
 \hypersetup{
    unicode=false,          
    pdftoolbar=true,        
    pdfmenubar=true,        
    pdffitwindow=false,     
    pdfstartview={FitH},    
    pdftitle={PhD Research Planning Report Template},    
    pdfauthor={Aldo Faisal},     
    pdfsubject={AI4Health Centre Template},   
    pdfcreator={Creator},   
    pdfproducer={Producer}, 
    pdfkeywords={keyword1, key2, key3}, 
    pdfnewwindow=true,      
    colorlinks=true,       
    linkcolor=red,          
    citecolor=green,        
    filecolor=magenta,      
    urlcolor=cyan           
}

\hypersetup{
 colorlinks=false,
 citecolor=blue,
 linkcolor=blue,
 urlcolor=blue}
\usepackage{float}
\usepackage{csquotes}
\usepackage[style=numeric,citestyle=numeric-comp,backend=biber,sorting=none]{biblatex}
\bibliography{Ethics_Paper}

\begin{document}
\begin{titlepage}

\newcommand{\HRule}{\rule{\linewidth}{0.5mm}} 
\setlength{\topmargin}{0in}
\center 

\HRule \\[0.5cm]
{ \huge Developing moral AI to support antimicrobial decision making}\\[0.5cm] 
\HRule \\[1cm]
 

\begin{minipage}{1\textwidth}
\begin{flushleft} \large
\raggedright
\emph{Authors:}
William J Bolton\textsuperscript{1, 2, 3}, Cosmin Badea\textsuperscript{3}, Pantelis Georgiou\textsuperscript{1, 4}, Alison Holmes\textsuperscript{1, 5, 6}, Timothy M Rawson\textsuperscript{1, 5} \\[0.5cm] 

\emph{Affiliations:}
\textsuperscript{1}Centre for Antimicrobial Optimisation, Imperial College London, UK.\textsuperscript{2}AI4Health Centre for Doctoral Training, Imperial College London, UK. \textsuperscript{3}Department of Computing, Imperial College London, UK.\textsuperscript{4}Centre for Bio-inspired Technology, Department of Electrical and Electronic Engineering, Imperial College London, UK.\textsuperscript{5}National Institute for Health Research, Health Protection Research Unit in Healthcare Associated Infections and Antimicrobial Resistance, Imperial College London, UK. \textsuperscript{6}Department of Infectious Diseases, Imperial College London, UK.\\[0.5cm]

\emph{Contact information:}
william.bolton@imperial.ac.uk, 
\\[0.5cm]
\end{flushleft}
\end{minipage}

\vfill 

\end{titlepage}

\newpage
\begin{abstract} 
Artificial intelligence (AI) assisting with antimicrobial prescribing raises significant moral questions. Utilising ethical frameworks alongside AI-driven systems, while considering infection specific complexities, can support moral decision making to tackle antimicrobial resistance.
\end{abstract}

\newpage
\section{AI antimicrobial decision making is morally complex}

Antimicrobials are drugs that kill or inhibit the growth of bacteria. The use or overuse of antimicrobials is a major driver of antimicrobial resistance (AMR) in humans \cite{holmes_understanding_2016}. AMR is now a leading global cause of mortality, killing more people than HIV or malaria in 2019 \cite{murray_global_2022}. To address AMR, a multi-modal approach is required that includes improving diagnosis, developing new antimicrobials, and critically preserving the effectiveness of our currently available agents. Recently, artificial intelligence (AI), and in particular machine learning (ML) and deep learning (DL) techniques, have shown great promise for providing decision support in healthcare, with models achieving accuracy on par and even above that of human experts \cite{topol_high-performance_2019, esteva_guide_2019}. Within the field of infection, models that support decision making have been developed to select appropriate antimicrobials and, more recently, predict COVID-19 outcomes \cite{rawson_systematic_2017, peiffer-smadja_machine_2020, wynants_prediction_2020}. Additionally, AI is now being applied in combination with diagnostic techniques to predict AMR development \cite{lv_review_2020}.\\

The development and adoption of AI-based decision support tools to support improved antimicrobial use raises significant moral questions. Most notably, antimicrobial prescribing decision support systems must, we argue, obtain a moral balance between the needs of an individual patient and those of wider and future society. Currently, when making decisions on whether or not to prescribe antimicrobials to a patient, clinicians evaluate the potential risks and rewards \cite{brink_best_2020, butler_antibiotics_2001}. Treating the patient may reduce suffering and disease progression, but risks driving the evolution of AMR as well as causing adverse effects \cite{holmes_understanding_2016, langford_is_2017}. Not prescribing antimicrobials has the opposite risk/reward profile. National and local guidelines are important in decision making as they help clinicians follow evidence-based practice and understand when specific drugs should be administered in certain situations. However, in our experience the antimicrobial decision making process is highly individualised and unfortunately frequently undertaken with limited information, meaning it is unclear which antimicrobial the infectious pathogen is susceptible to, or even if the patient could recover without treatment. This often results in difficult decisions that are not black-or-white, the consequences of which may or may not help the patient and may or may not harm future populations. As such, the morally right decision is often unclear. Incorporating such concepts into AI systems is complex but may be supported by developing a consensus on the optimal approach to decision making in this context. Within this text we aim to explore potential ethical frameworks and nuances that may be applied to define what is ethical or not during the development of AI-based clinical decision support systems (CDSSs) for antimicrobial optimisation.\\



\section{AI and ethical frameworks to support moral decision making}


Taking the perspective of a utilitarian ethicist, who believes that an action is only 'good' if it creates utility (frequently measured as happiness) \cite{sep-consequentialism}, a moral balance may be achieved and applied to the development of AI-driven CDSSs. Utilitarianism in healthcare can be evaluated using multiple techniques. These include total, average, minimum and total-average utility \cite{sep-consequentialism}. Maximising total-average utility is likely to be of greatest importance in the context of AMR and healthcare, given it aims to optimise the average happiness for those people who are currently alive \cite{sep-consequentialism}. This aligns with the UK's General Medical Council (GMC) duties of a doctor where the objective is to maximise that patient's health and extend life, while also considering wider society and providing equality \cite{gmc_duties_2019}. Frameworks such as Bentham's felicific calculus are commonly used within utilitarianism and can be applied to antimicrobial decision making processes to quantify the utility of an action. Table. \ref{fig:felicific_calculus} provides an illustration of this calculus and its application to the decision of starting antimicrobial treatment. When taken as a whole this framework suggests that to justify prescribing antimicrobials, the intensity and duration utility gained from the individual patient must outweigh the negative impact on everyone else. The only utilitarian scenario when this could occur is with a utility monster, who gains significantly greater sums of utility from actions. This seems improbable in our reality and to maximise total-average happiness we must look to maintain all life \cite{post_breaking_2022}. Importantly though, certainty and propinquity can't be defined without more information. As such AI models can contribute towards utility evaluations through estimating the impact of the selected agent on the development of AMR versus the likelihood of clinical efficacy. By combining moral frameworks and AI-based CDSSs we can ensure the utility of an action can be quantified and understand potential individual and societal implications. As discussed above, decision making in antimicrobial prescribing is frequent and both morally and technically complex. Ensuring decisions are morally justified alongside utilising the analytical prowess of AI systems would thus support decision making and help address AMR by reducing inappropriate antimicrobial prescribing.\\

\begin{table}[H]
\centering
\caption{\footnotesize Overview of Bentham's felicific calculus variables and example application to starting antimicrobial treatment.}
\includegraphics[width=\linewidth]{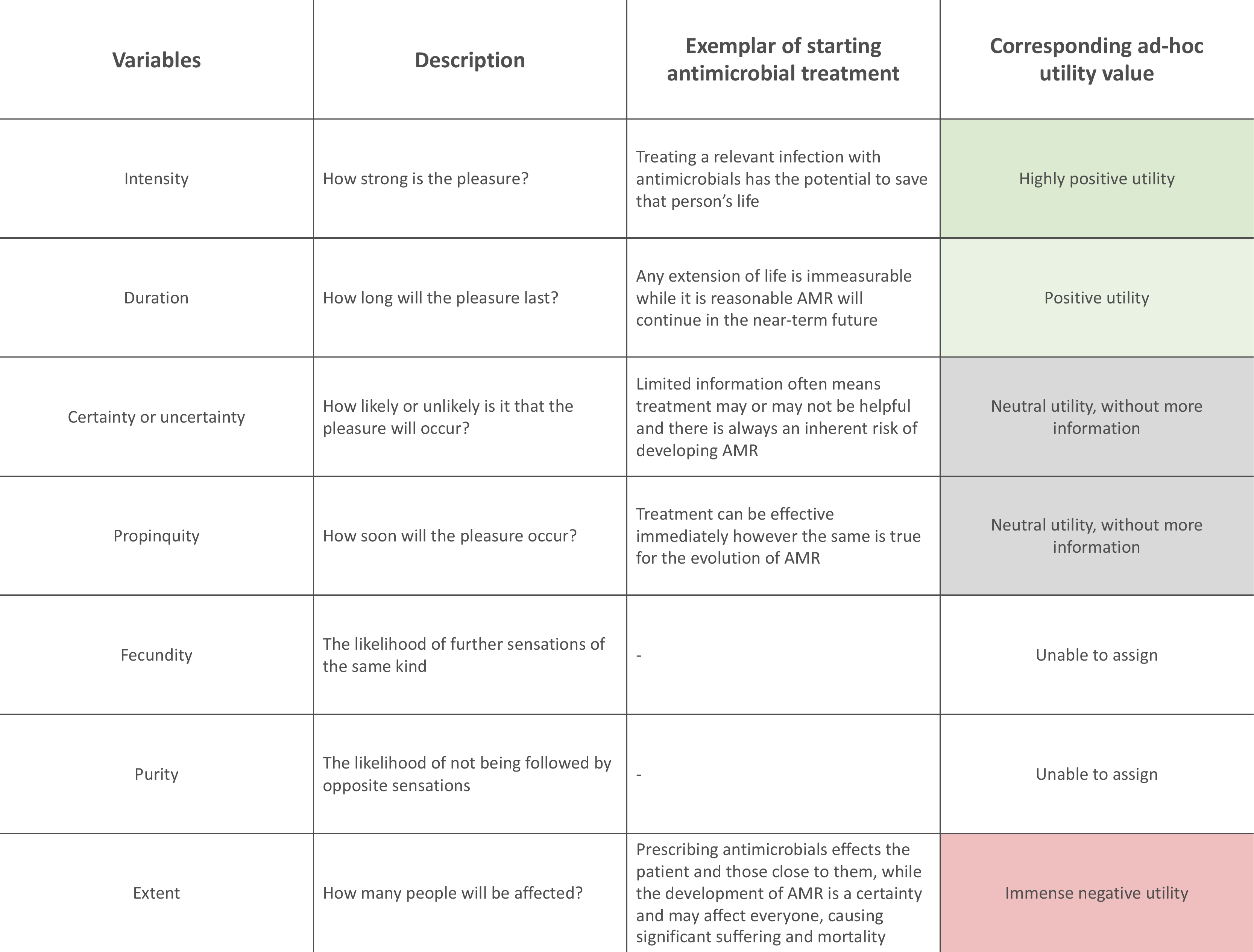}
\label{fig:felicific_calculus}
\end{table}

\newpage
Alternatively, deontological or virtue-based frameworks may be explored. Deontology applies duty-based ethics to determine if an action is morally good or bad based on whether one acts in accordance with our duty \cite{sep-ethics-deontological}. Duty-based ethics is very amenable to rule-based decision making, as such one could attempt to figure out what the perfect and imperfect duties are in this situation, based on, for instance, Kant's categorical imperative \cite{sep-kant-moral}. The UK's GMC 'duties of a doctor' or the Hippocratic oath can be considered a more deontology focused approach given care of the patient is of primary concern \cite{gmc_duties_2019}. Although these principals were designed to be generalist and are not tailored for the significant ethical dilemma posed by antimicrobial prescribing. Virtue ethics on the other hand focuses on the moral character of the person carrying out actions, an assessment of which can be made based on comparisons with a virtuous person who possesses and embodies the virtues \cite{sep-ethics-virtue}. In this context, virtues would need to be defined and the question of 'What would a virtuous person do?' considered. One could argue that virtuous clinicians may act as moral exemplars for complex decision making \cite{hindocha_moral_2021}, but common moral dilemmas arise in the context of decision making to address AMR. For example, weighing up the potential number of lives lost and taking action to reduce the number of deaths versus having personal responsibility for an individual's outcome \cite{post_breaking_2022}, and whether acting under a high pressure situation is the same as not acting. By comparing other ethical frameworks to a utilitarian perspective, we can insinuate that different ethical stances may have contrasting viewpoints on what is considered morally right with regards to prescribing antimicrobials. Therefore, depending on the developers, regulators and users stance, different frameworks and levels of tolerance could be considered within AI-driven CDSSs.\\

\section{Technological and clinical considerations}

Developing moral AI-driven CDSSs that incorporate appropriate ethical frameworks may support the wider adoptions of such systems. However further technological and infection specific clinical factors need to be considered \cite{lysaght_ai-assisted_2019, grote_ethics_2020}. To start with, how should information be combined to reach a morally good decision? Important considerations required for decision making are unlikely to be processed through one model. For example, one system may output the anticipated antimicrobial risk/ reward profile to the patient while another produces a prediction for the likelihood of AMR development. As such, some sort of aggregation model or function will be required \cite{badea_have_2021}, which gets particularly complex when patients' preferences are taken into account as is the case with shared decision making \cite{butler_antibiotics_2001}. This leads on to the question of who is accountable for AI assisted decisions. What happens if a clinician ignores or disagrees with the CDSS, or if recommendations lead to unexpected negative consequences, on either the patient or the public side? Currently, clinicians are responsible for their actions and in case of dispute must provide a rational and justification. Similar accountability is likely to be required for the developers of such AI-based systems, particularly in areas of high consequence and moral dispute such as infectious diseases. This brings up the issue of transparency. Complex deep learning models are often considered black boxes with limited explainability and interpretability. A significant amount of research is trying to tackle this problem \cite{hindocha_moral_2021, barredo_arrieta_explainable_2020}, but what is acceptable for an antimicrobial prescribing decision is yet to be fully defined. Finally, a model is only as good as the data used to train it, and as such large, diverse and representative datasets must be used to prevent harm and inequality for minority and disadvantaged groups. This is particularly apparent for infectious diseases where people of different ethnic/racial backgrounds have different infection-related risks and outcomes, and research has shown a strong association between poor socioeconomic status, increased rates of infection and AMR \cite{ecdc_health_2013, alividza_investigating_2018}. This has been emphasised during the COVID-19 pandemic where patients from Black, Asian and minority ethnic (BAME) groups or individuals with co-morbidities have seen significantly worse treatment outcomes than those from a white background or with no underlying health conditions \cite{wynants_prediction_2020}. A large number of AI-based CDSSs were developed to assist with COVID-19 however most models suffered from significant biases and lack of generalisation \cite{wynants_prediction_2020}. These issues highlight that due to the complex and regulated nature of healthcare, additional factors must be considered when evaluating the morality of AI-based CDSSs.\\

As AI-driven CDSSs are developed to optimise antimicrobial use, greater focus on how these algorithms are trained and developed in an ethical and fair manner is required. Considering antimicrobial prescribing, expert opinion often deviates depending on the specific scenario, while relying on a single methodology such as an antibiogram can be unreliable or not correlate with expected response. As such when tackling clinically uncertain problems, relying on the latest gold standard information from a more holistic and local perspective, as well as employing powerful ML techniques to uncover unknown relationships can be perceived as a favourable approach. Ultimately the aim of an AI-driven CDSS for antimicrobial prescribing is to drive behaviour change in a way that optimises treatment outcome, whilst minimising the development of side effects or AMR. Such systems will need to provide prescribers with a high confidence in their prescribing recommendations for a given patient so they can fulfill their 'duties of a doctor', while also incorporating, for example, a utilitarian society benefit. How to best present this information to a clinician and drive a change in prescribing behaviour through an AI tool is an important question. The co-design of systems has been proposed to assist CDSS adoption and transparency \cite{rawson_artificial_2019}. While carrying-out AI literacy training, understanding what behaviour change is needed and how the tool fits into the clinical prescribing workflow are vital to ensure impact and appropriate patient and doctor engagement. For infectious diseases specifically, additional factors must be investigated. The evolutionary process of pathogenic microorganisms and thus the development of AMR occurs on a human life timescale. Hence, when tackling AMR, AI should be temporally dynamic so that it is sensitive to microbiological evolutionary. In addition, systems should be geographically revised given infectious diseases, resistance rates, and antimicrobial availability vary dramatically by region \cite{holmes_understanding_2016}. Moreover heterogeneity is needed with regards to antimicrobial prescribing, because uniform treatment drives AMR \cite{holmes_understanding_2016}. These factors increase the importance of algorithmic transparency and accessibility of live local medical data. Applying AI to support optimal antimicrobial prescribing is therefore a complex but a crucial endeavor to try and counteract AMR.\\


\section{Conclusion}
Moral frameworks have been designed to help humans make ethical decisions. As our species advances into a new age with AI we must draw from these principals and adapt them as appropriate, to enable 'intelligent' decision systems to tackle global problems, such as the emerging AMR crisis, in a moral way. The development of AI-driven CDSSs for antimicrobial prescribing should help to tackle AMR, but such systems must additionally conform to the specific requirements of modern infectious diseases medicine.\\


\newpage
\addcontentsline{toc}{section}{References}
\footnotesize
\printbibliography
\newpage
\end{document}